\documentstyle[preprint,aps,prb]{revtex}
\begin{document}
%\preprint{ UCD-}
\draft
\title{
Iterated perturbation theory for the attractive Holstein and Hubbard models
}
\author{J. K. Freericks }
\address{
Department of Physics, University of California, Davis, CA 95616}
\author{ Mark Jarrell}
\address{Department of Physics, University of Cincinnati, Cincinnati, OH 45221}
\date{\today}
\maketitle
\widetext
\begin{abstract}
A strictly truncated (weak-coupling) perturbation theory is applied to the 
attractive Holstein and Hubbard models in infinite dimensions.  These results
are qualified by comparison with essentially exact Monte Carlo results.
The second order
iterated perturbation theory is shown to be quite accurate in calculating
transition temperatures for retarded interactions, but is not as accurate
for the self energy or the irreducible vertex functions themselves.  Iterated
perturbation theory is carried out thru fourth order for the Hubbard model.  
The self energy is quite accurately reproduced by the theory, but the vertex
functions are not.  Anomalous behavior occurs near half filling because the
iterated perturbation theory is not a conserving approximation.
\end{abstract}
\section{Introduction}

Recently Metzner and Vollhardt\cite{metzner_vollhardt} showed that the 
fermionic many-body problem simplifies in the limit of infinite spatial
dimensions.  The $d\rightarrow\infty$ limit is taken in such a fashion
that the dynamics are local in space, and the lattice problem can be mapped
onto a self-consistently embedded Anderson impurity model.  The self energy
and irreducible vertex functions of the impurity problem are employed to
determine the Green's functions and susceptibilities of the infinite-dimensional
lattice.  This many-body problem can essentially be solved exactly by using
the quantum Monte Carlo (QMC) techniques of Hirsch and Fye\cite{hirsch_fye}
to extract the self energy and irreducible vertex functions of the
relevant impurity problem.  A variety of models have already been examined in
this fashion: the Hubbard 
model\cite{jarrell,kotliar_hubb,georges_hubb,vollhardt_hubb}; periodic Anderson
model\cite{jarrell_anderson}; and the Holstein 
model\cite{freericks_jarrell_scalapino}.

The infinite-dimensional limit provides a unique testing ground for various
approximation techniques, since one can compare them to the benchmark QMC
results.  Previous work has concentrated on the strong-coupling 
limit\cite{freericks_strong,vollhardt_strong}
(perturbation theory in the kinetic energy) which can be shown to be an
excellent approximation in the region of moderate to strong coupling.

Calculations in the weak-coupling limit are not under as good control.  The 
main emphasis so far has been on conserving 
approximations\cite{muellerhartmann,hirashima,feinberg,freericks_conserving}
(such as the fluctuation-exchange approximation) to the many-body problem.
Conserving approximations are popular because they
satisfy the requisite conservation laws for electronic
charge, energy, and momentum\cite{baym}.  It turns out, however, that 
conserving approximations suffer from some serious flaws: (1) truncated
approximations (thru fourth order)
do not show a turnover in $T_c$ as the strong-coupling regime
is approached; (2) infinite summations of certain classes of diagrams
can produce a turnover in $T_c$, but the self energy is overestimated,
and many-body features (such as
the upper and lower Hubbard bands) do not appear in the single-particle
spectral function.  Similar deficiencies where found by White
in his examination of the Anderson impurity model itself\cite{white}.

An alternate approximation scheme, that does not suffer from the above problems
(but is not conserving), has been introduced by Georges and 
Kotliar\cite{georges_kotliar} and is known as iterated perturbation theory
(IPT).  In this case, the perturbation theory is strictly truncated
at a finite order (as opposed to a conserving approximation that sums
the infinite class of diagrams corresponding to self energy insertions into
the dressed Green's function).  The motivation for this approximation comes
from the pioneering work of Yosida and Yamada\cite{yosida_yamada}, who found
that truncated perturbation theory was an extremely accurate approximation
for the Anderson impurity model because the sum total of all classes of diagrams
at a given order is an order of magnitude smaller than any of its constituent
parts.  This remarkable near cancellation of diagrams was further verified
to high order by Zlati\'c and Horvati\'c\cite{horvatic_zlatic_bethe} who 
found that
the exact Bethe ansatz solutions can be rearranged into a power series whose
coefficients die off rapidly.

Horvati\'c and Zlati\'c\cite{horvatic_zlatic_2nd}
generalized Yosida and Yamada's
analysis off of half filling and found that the strictly truncated
perturbation series was also accurate in the doped case.  However, anomalous 
behavior begins to enter if the coupling strength is increased too far
and the electron concentration is close to half filling.  Ferrer, 
Mart\'in-Rodero, and Flores\cite{ferrer_martinrodero_flores}
realized that the reason why the perturbation theory fails
more rapidly away from half filling is because the second-order perturbation
theory correctly reproduces the atomic limit at half-filling, but not
away from half filling.  They proposed an {\it ad hoc} interpolation scheme
that properly reproduces the atomic limit everywhere and removes some of the
anomalous behavior of the Horvati\'c and Zlati\'c approximation.  An alternate 
interpolation scheme has been proposed by Neal\cite{neal} based upon functional
integral techniques which appears to be quite accurate for all values of 
the interaction strength but is restricted to half filling.

In this contribution, we will examine the iterated perturbation theory through
second order for the Holstein model and through fourth order for the negative-U 
Hubbard model.  Comparison of the self energy, the irreducible vertex functions, and the transition temperatures will be made to the exact QMC solutions.  In
Section II the infinite-dimensional formalism is reviewed along with the
details of the IPT approximation.  Section III contains the results for the
Holstein model, while Section IV holds the results for the Hubbard model.
Conclusions and discussion follow in Section V.

\section{Formalism}

The electron-phonon interaction is generally believed to be responsible for the
superconductivity in most low-temperature superconductors\cite{parks,reviews}.
The conventional theory used to describe superconductivity was introduced
by Migdal\cite{migdal} and Eliashberg\cite{eliashberg}.  Migdal-Eliashberg
(ME) theory ignores vertex corrections which can be shown to be proportional
to powers of the Debye frequency divided by the Fermi energy, and hence are 
small corrections for most real materials.

Here, iterated perturbation theory is examined in a systematic fashion for the
attractive electron-phonon interaction described by the
Holstein\cite{holstein} model
(the effects of Coulomb repulsion are explicitly neglected).
The Holstein model consists of conduction electrons  that
interact with localized (Einstein) phonons:
\begin{equation}
  H = - {{t^*}\over{2\sqrt{d}}} \sum_{\langle j,k\rangle \sigma} ( 
   c_{j\sigma}^{\dag }
  c_{k\sigma} + c_{k\sigma}^{\dag  } c_{j\sigma} ) + \sum_j  (gx_j-\mu)
  (n_{j\uparrow}+n_{j\downarrow}-1) + {{1}\over{2}} M \Omega^2 \sum_j x_{j}^2 +
  {{1}\over{2}} \sum_j {p_{j}^2\over M}
\label{eq: holham}
\end{equation}
where $c_{j\sigma}^{\dag}$ ($c_{j\sigma}$) creates (destroys) an electron at
site $j$ with spin $\sigma$, $n_{j\sigma}=c_{j\sigma}^{\dag}c_{j\sigma}$ is
the electron number operator, and $x_j$ ($p_j$) is the phonon coordinate
(momentum) at site $j$.  The hopping matrix elements connect the nearest
neighbors of a hypercubic lattice in $d$-dimensions.  The 
unit of energy is chosen to be the rescaled matrix
element $t^*$.  The phonon has a mass $M$ (chosen to
be $M=1$), a
frequency $\Omega$, and a spring constant $\kappa\equiv M\Omega^2$ associated
with it.  The electron-phonon coupling constant (deformation potential)
is denoted by $g$; the
effective electron-electron interaction strength is then the bipolaron binding energy
\begin{equation}
  U\equiv - {{g^2}\over{M\Omega^2}}=-{g^2\over\kappa} \quad .  
 \label{eq: udef}
\end{equation}
The chemical potential is denoted by $\mu$ and particle-hole symmetry occurs
for $\mu=0$.

In the instantaneous limit where $U$ remains finite and $g$ and $\Omega$ are
large compared to the bandwidth $(g,\Omega\rightarrow\infty ,U={\rm
finite})$, the Holstein model maps onto the attractive Hubbard 
model\cite{hubbard}
\begin{equation}
  H = - {{t^*}\over{2\sqrt{d}}} \sum_{\langle j,k\rangle \sigma} ( 
   c_{j\sigma}^{\dag }
  c_{k\sigma} + c_{k\sigma}^{\dag  } c_{j\sigma} ) -\mu\sum_j (n_{j\uparrow}+
n_{j\downarrow}) + U \sum_j   (n_{j\uparrow}-{1\over 2})(n_{j\downarrow}-
{1\over 2})
\label{eq: hubbham}
\end{equation}
with $U$ defined by Eq.~(\ref{eq: udef}).

The infinite-dimensional limit of Metzner and Vollhardt\cite{metzner_vollhardt}
is taken $(d\rightarrow\infty)$, in which the electronic many-body problem
becomes a local (impurity) problem that retains its complicated dynamics in
time.  The hopping integral is scaled to zero in such a fashion that the
free-electron kinetic energy remains finite while the self energy
for the single-particle Green's function and the irreducible vertex functions
have no momentum dependence and are functionals of the
local Green's function\cite{metzner_vollhardt,schweitzer_czycholl,metzner}.
This limit retains the strong-correlation effects that arise from
trying to simultaneously minimize both the kinetic energy 
and the potential energy.

The many-body problem is solved by mapping it onto an auxiliary impurity
problem\cite{brandt_mielsch,okhawa} in a time-dependent field 
that mimics the hopping of an electron onto
a site at time $\tau$ and off the site at a time $\tau '$.  The 
action for the impurity problem is found by integrating out the 
degrees of freedom associated with other lattice sites in a path-integral 
formalism.\cite{georges_kotliar}  The result is an effective action
\begin{eqnarray}
S_{eff.} &= \sum_{\sigma} \int_0^{\beta} d\tau \int_0^{\beta} d\tau '
c_{\sigma}^{\dag}(\tau)G_0^{-1}(\tau-\tau ')c_{\sigma}(\tau ')
+\sum_{\sigma}\int_0^{\beta}d\tau[gx(\tau)-\mu][c_{\sigma}^{\dag}(\tau)
c_{\sigma}(\tau)-{1\over 2}]\cr
&+ {1\over2}M\int_0^{\beta}d\tau [\Omega^2x^2(\tau)+\dot x^2(\tau)]
\label{eq: seff}
\end{eqnarray}
where $G_0^{-1}$ is the ``bare'' Green's function that contains
{\it all of the dynamical information of the other sites of the lattice}.
The interacting Green's function, defined to be
\begin{equation}
G(i\omega_n)\equiv -\int_0^{\beta} d\tau e^{i\omega_n\tau}
{{\rm Tr} \langle e^{-\beta H}T_{\tau} c(\tau)
c^{\dag}(0)\rangle\over {\rm Tr} \langle e^{-\beta H}\rangle } ~,
\label{eq: greendef}
\end{equation}
is determined by Dyson's equation
\begin{equation}
G_n^{-1}\equiv G^{-1}(i\omega_n) = G_0^{-1}(i\omega_n)-\Sigma (i\omega_n).
\label{eq: gdef}
\end{equation}

A self-consistency relation is required in order to determine the bare
Green's function $G_0$.  This is achieved by mapping the impurity problem 
onto the infinite-dimensional lattice thereby equating the full Green's 
function for the impurity problem with the local Green's function for 
the lattice
\begin{equation}
G_{jj}(i\omega_n)=\sum_{\bf k} G({\bf k},i\omega_n) = \sum_{\bf k}
[i\omega_n+\mu-E({\bf k})-\Sigma(i\omega_n)]^{-1}
= F_{\infty}[i\omega_n+\mu -\Sigma(i\omega_n)].
\label{eq: gloc}
\end{equation}
Here $F_{\infty}(z)$ is the scaled complimentary error function of a complex 
argument
\begin{equation}
F_{\infty}(z)\equiv {1\over{\sqrt{\pi}}}\int_{-\infty}^{\infty} dy 
{\exp(-y^2)\over{z-y}} =-i{\rm sgn}[{\rm Im}(z)]\sqrt{\pi}e^{-z^2}{\rm erfc}
\{-i{\rm sgn}[{\rm Im}(z)]z\}. 
\label{eq: fdef}
\end{equation}
The dynamics of the (local) impurity problem are identical to the dynamics
of the Anderson impurity 
model\cite{schweitzer_czycholl,brandt_mielsch,okhawa,georges_kotliar,jarrell}.
This many-body problem can either be solved exactly with the QMC algorithm
of Hirsch and Fye\cite{hirsch_fye}, or it can be approximately solved
by employing a truncated perturbative
expansion.  The impurity is self-consistently embedded in the host, since
it must satisfy the self-consistency relation in Eq.~(\ref{eq: gloc}).

It is important to note that since one does not {\it a priori} know the 
bare Green's function $G_0^{-1}$ in Eq.~(\ref{eq: seff}), one must iterate 
to determine a self-consistent solution for the Green's
function of the infinite-dimensional lattice.  This is achieved by calculating
the self energy as a functional of the bare Green's function $G_0$, and 
then determining the new local Green's 
function from the approximate self energy and
Eq.~(\ref{eq: gloc}).  The new bare Green's function is then calculated
from the Dyson equation in Eq.~(\ref{eq: gdef}). This process is iterated until 
convergence is achieved [for the perturbation theory approximation
the maximum variation of each $G(i\omega_n)$ is less than one part in
$10^8$ which typically takes between 5 and 30 iterations, for the QMC 
calculations the algorithm is iterated from 7 to 9 times].  Note that the
iteration required to determine a self-consistent bare Green's function 
$G_0$ should 
not be confused with the the iterative techniques used to self-consistently
sum all of the self-energy insertions in a self-consistent perturbation 
theory---{\it the perturbation series is always strictly truncated at a finite
order here}.

The QMC algorithm proceeds by discretizing the imaginary-time
interval from $0$ to $\beta$ into $L$ time slices of equal width
$\Delta\tau=\beta/L$ and evaluating the relevant path integrals 
in a grand canonical scheme.  Both local moves, in which the phonon
coordinate is shifted by a different amount at each time slice, and
global moves, in which the phonon coordinate
is shifted by a uniform amount for every time slice, are incorporated
into the QMC algorithm\cite{rhein1}.
%% JIM perhaps we should also reference the Georgia conference article here?
The values of $L$ used ranged from 20
to 160 with the largest values of $\Delta\tau$ reserved for the lowest
temperatures (usually $\Delta\tau $ was fixed at 0.4).  
No sign problem was found at any filling.

Static two-particle properties are also easily calculated since the 
irreducible vertex function is local\cite{zlatic_horvatic_infd}.  The 
static susceptibility for CDW order is given by
\begin{eqnarray}
\chi^{CDW}({\bf q})&\equiv& {1\over2N}\sum_{{\bf R}_j-{\bf R}_k\sigma\sigma'}
e^{i{\bf q}\cdot
({\bf R}_j-{\bf R}_k)} T\int_{0}^{\beta} d\tau \int_{0}^{\beta} d\tau '
[\langle n_{j\sigma}(\tau) n_{k\sigma'}(\tau ')\rangle - 
\langle n_{j\sigma}(\tau)\rangle\langle n_{k\sigma'}
(\tau ')\rangle ] \cr
&\equiv& T\sum_{mn} \tilde\chi^{CDW} ({\bf q},i\omega_m,i\omega_n)=
T\sum_{mn} \tilde\chi_{mn}^{CDW} ({\bf q})\quad,
\label{eq: chicdw}
\end{eqnarray}
at each ordering wavevector ${\bf q}$.  Dyson's equation for the two-particle
Green's function becomes\cite{jarrell,zlatic_horvatic_infd}
\begin{equation}
\tilde\chi_{mn}^{CDW}({\bf q})=\tilde\chi_{m}^0({\bf q})\delta_{mn}
-T\sum_p \tilde\chi_m^0({\bf q})\Gamma_{mp}^{CDW}\tilde
\chi_{pn}^{CDW}({\bf q})\quad ,
\label{eq: cdwdys}
\end{equation}
with $\Gamma_{mn}^{CDW}$ the (local) irreducible vertex function in the CDW
channel.  

The
bare CDW susceptibility $\tilde\chi_n^0({\bf q})$ in 
Eq.~(\ref{eq: cdwdys}) is defined in terms of the {\it dressed}
single-particle Green's function
\begin{eqnarray}
\tilde\chi_n^0({\bf q})&\equiv&-{1\over N} \sum_{\bf k} G_n({\bf
k})G_n({\bf k+q})\cr
&=&-{1\over{\sqrt{\pi}}}{1\over{\sqrt{1-X^2({\bf q})}}}\int_{-\infty}^{\infty}
dy {{e^{-y^2}}\over{i\omega_n+\mu-\Sigma_n-y}}F_{\infty}\left [ {{i\omega_n+
\mu-\Sigma_n-X({\bf q})y}\over{\sqrt{1-X^2({\bf q})}}}\right ]
\label{eq: chi0cdw}
\end{eqnarray}
and all of the wavevector dependence is included in the
scalar\cite{brandt_mielsch,muellerhartmann} $X({\bf q})
\equiv \sum\nolimits_{j=1}^d \cos {\bf q}_j/d$.  The mapping ${\bf q}
\mapsto X({\bf q})$ is a many-to-one mapping that determines an equivalence
class of wavevectors in the Brillouin zone.  ``General'' wavevectors are
all mapped to $X=0$ since $\cos {\bf q}_j$ can be thought of as a random
number between $-1$ and 1 for ``general'' points in the Brillouin zone.
Furthermore, all possible values of $X$ $(-1\le X\le 1)$ can be labeled
by a wavevector that lies on the diagonal of the first Brillouin zone extending
from the zone center $(X=1)$ to the zone corner $(X=-1)$.  The presence
of incommensurate order in the attractive Holstein model is restricted to a very
narrow region of parameter space\cite{freericks_jarrell_scalapino,feinberg}
so only the ``antiferromagnetic'' point $X=-1$ is considered for CDW order.
The integral for $\tilde\chi_m^0(X)$
in Eq.~(\ref{eq: chi0cdw}) can then be
performed analytically\cite{brandt_mielsch}
$\tilde\chi_n^0(X=-1)=-{{G_n}/({i\omega_n+\mu-\Sigma_n}})$.
The irreducible vertex function $\Gamma_{mn}^{CDW}$ is either directly
calculated in a perturbative expansion (IPT) or
is determined by inverting the Dyson equation in Eq.~(\ref{eq: cdwdys}) (QMC).

A similar procedure is used for the singlet $s$-wave SC channel.  The
corresponding definitions are as follows:  The static susceptibility
in the superconducting channel is defined to be
\begin{eqnarray}
\chi^{SC}({\bf q})&\equiv& {1\over N}\sum_{{\bf R}_j-{\bf R}_k}
e^{i{\bf q}\cdot
({\bf R}_j-{\bf R}_k)} T\int_{0}^{\beta} d\tau \int_{0}^{\beta} d\tau '
\langle c_{j\uparrow}(\tau)
c_{j\downarrow}(\tau)c_{k\downarrow}^{\dag}(\tau ')c_{k\uparrow}^{\dag}(\tau ')
\rangle  \cr
&\equiv& T\sum_{mn} \tilde\chi^{SC} ({\bf q},i\omega_m,i\omega_n)=
T\sum_{mn} \tilde\chi_{mn}^{SC} ({\bf q})\quad,
\label{eq: chisc}
\end{eqnarray}
for superconducting pairs that carry momentum ${\bf q}$;  Dyson's equation
becomes
\begin{equation}
\tilde\chi_{mn}^{SC}({\bf q})=\tilde\chi_{m}^0{'}({\bf q})\delta_{mn}
-T\sum_p \tilde\chi_m^0{'}({\bf q})\Gamma_{mp}^{SC}\tilde
\chi_{pn}^{SC}({\bf q})\quad ,
\label{eq: scdys}
\end{equation}
with $\Gamma_{mn}^{SC}$ the corresponding irreducible vertex function for the 
SC channel; the bare pair-field susceptibility becomes
\begin{eqnarray}
\tilde\chi_n^0{'}({\bf q})&\equiv&{1\over N} \sum_{\bf k} G_n({\bf
k})G_{-n-1}({\bf -k+q})\cr
&=&{1\over{\sqrt{\pi}}}{1\over{\sqrt{1-X^2({\bf q})}}}\int_{-\infty}^{\infty}
dy {{e^{-y^2}}\over{i\omega_n+\mu-\Sigma_n-y}}F_{\infty}\left [ {{i\omega_{-n-1}+
\mu-\Sigma_{-n-1}-X({\bf q})y}\over{\sqrt{1-X^2({\bf q})}}}\right ]
\label{eq: chi0sc}
\end{eqnarray}
with the special value 
$\tilde\chi_n^0{'}(X=1)=-{\rm Im}G_n/{\rm Im}(i\omega_{n}-\Sigma_{n})$ for the
SC pair that carries no net momentum; and finally the irreducible vertex 
function is also  either directly
calculated in a perturbative expansion (IPT) or
is determined by inverting the Dyson equation in Eq.~(\ref{eq: scdys}) (QMC).

There are two different approximation schemes that can be used for determining
the irreducible vertex functions in a perturbative expansion.  The first scheme,
denoted IPT, is in the spirit of Yosida and Yamada's original 
analysis\cite{yosida_yamada}: the
vertex functions are also expanded in a perturbation series that is strictly 
truncated at a finite order.  In the second scheme, denoted IPT$^*$, the
fully dressed local Green's functions are used to calculate the irreducible 
vertex functions.  In general, one might expect the IPT approximation to be
more accurate than the IPT$^*$ approximation from the work of Yosida and
Yamada\cite{yosida_yamada}, but the irreducible vertex functions were not 
investigated in their analysis of the Anderson impurity model, so it is not
{\it a priori} clear which will be a better approximation.

The transition temperature of the infinite-dimensional Holstein 
model is now found by calculating the temperature at which the relevant 
susceptibility diverges (CDW or SC).
One can determine this transition temperature by
finding the temperature where the scattering matrix (in the relevant
channel)
\begin{equation}
T_{mn}=-T\Gamma_{mn}\chi_n^0\quad,
\label{eq: Tmat}
\end{equation}
has unit eigenvalue\cite{owen_scalapino}.  Note that the fully dressed
Green's functions are always used in calculating the bare susceptibility 
$\chi^0$
because the bare susceptibility is a property of the infinite-dimensional
lattice, and not the Anderson impurity problem, and we expect to find better
agreement with the QMC results if we use the correct bare susceptibility.

\section{Holstein Model}

The diagrammatic expansion for the IPT is depicted in Figures 1 and 2.
The wiggly lines denote the bare phonon propagator, and the straight lines 
denote the bare Green's function $G_0$.  Figure 1 shows the self energy for the 
IPT approximation through second order.  The self energy includes,
respectively, the Hartree term (which is a constant that is reabsorbed into
the chemical potential), the Fock term, the second-order term that dresses
the phonon propagator, the lowest-order vertex correction, and the second-order
rainbow diagram (which corresponds to the self-energy insertion of the 
first-order Fock diagram into the first-order Fock diagram). 

To be more explicit, the self energy for the second-order IPT approximation is
\begin{eqnarray}
\Sigma_n&=&-UT\sum_r{\Omega^2\over \Omega^2+\nu_{n-r}^2} G_0(i\omega_r)\cr
&+&U^2T^2\sum_{rs}[-2{\Omega^2\over\Omega^2+\nu_{n-r}^2}+{\Omega^2\over\Omega^2
+\nu_{r-s}^2}]{\Omega^2\over\Omega^2+\nu_{n-r}^2}G_0(i\omega_r)G_0(i\omega_s)
G_0(i\omega_{n-r+s})\cr
&+&U^2T^2\sum_{rs}{\Omega^2\over\Omega^2+\nu_{n-r}^2}{\Omega^2\over\Omega^2
+\nu_{r-s}^2}G_0^2(i\omega_r)G_0(i\omega_s)
\label{eq: sigma_ipt}
\end{eqnarray}
which includes the Fock diagram contribution and the three second-order
contributions of Figure 1.  The bosonic Matsubara frequency $\nu_l$
is defined to be $\nu_l\equiv 2l\pi T$.

The self-consistency step involves first determining a new local Green's 
function
$G_n$ from the integral relation in Eq.~(\ref{eq: gloc}) with the
approximate self-energy of Eq.~(\ref{eq: sigma_ipt}).  Next the new bare
Green's function is found from the Dyson equation in Eq.~(\ref{eq: gdef})
using the same approximate self energy.  A new self energy is then calculated
from the new bare Green's function
[using Eq.~(\ref{eq: sigma_ipt})], and the process is iterated.
This iteration process terminates when the maximum deviation in the 
local Green's function is less than one part in $10^8$.

Once the Green's functions and self energies have been determined, the
irreducible vertex functions can be calculated for the CDW or SC
channels.  In the CDW channel [see Figure 2 (a)] one must
include both direct and exchange diagrams as well as the vertex corrections.
The result is
\begin{eqnarray}
\Gamma_{mn}^{CDW}&=&2U-2U^2T\sum_r[G_0(i\omega_r)G_0(i\omega_{m-n+r})
+G_0(i\omega_r)G_0(i\omega_{m+n-r})]
[{\Omega^2\over \Omega^2+\nu_{n-r}^2}]^2\cr
&-&U{\Omega^2\over \Omega^2+\nu_{m-n}^2}-2U^2T\sum_rG_0(i\omega_r)
G_0(i\omega_{m-n+r})
{\Omega^2\over \Omega^2+\nu_{m-n}^2}[{\Omega^2\over \Omega^2+\nu_{m-n}^2}
-{\Omega^2\over \Omega^2+\nu_{n-r}^2}]\cr
&+&U^2T\sum_rG_0(i\omega_r)G_0(i\omega_{m+n-r})
{\Omega^2\over \Omega^2+\nu_{m-r}^2}{\Omega^2\over\Omega^2+\nu_{n-r}^2}\quad ,
\label{eq: cdw_ipt}
\end{eqnarray}
for the IPT approximation.  The IPT$^*$ approximation has the same functional
form as in Eq.~(\ref{eq: cdw_ipt}), but the bare Green's function $G_0$ is
replaced by the dressed Green's function $G$ (this functional form is the
same as the one used for the conserving 
approximation\cite{freericks_conserving}).
Note that the vertex corrections (arising from the first-order exchange 
diagrams) modify the {\it interaction} in the CDW channel so that it properly
interpolates between the zero frequency limit $\Gamma^{CDW}\rightarrow 2U$
and the infinite frequency limit $\Gamma^{CDW}\rightarrow U$.  At an 
intermediate
frequency, the CDW interaction strength has a complicated temperature
dependence.  In the SC channel [see Figure 2 (b)] one finds
\begin{eqnarray}
\Gamma_{mn}^{SC}&=&U {\Omega^2\over \Omega^2+\nu_{m-n}^2}\cr
&+&U^2T\sum_r
G_0(i\omega_r)G_0(i\omega_{m-n+r})
{\Omega^2\over \Omega^2+\nu_{m-n}^2}[2{\Omega^2\over\Omega^2
+\nu_{m-n}^2}-{\Omega^2\over\Omega^2+\nu_{n-r}^2}-
{\Omega^2\over\Omega^2+\nu_{m+r+1}^2}]\cr
&-&U^2T\sum_rG_0(i\omega_r)G_0(i\omega_{m+n+r+1})
{\Omega^2\over\Omega^2+\nu_{m+r+1}^2}
{\Omega^2\over\Omega^2+\nu_{n+r+1}^2}\quad ,
\label{eq: sc_ipt}
\end{eqnarray}
for the IPT approximation.  Once again, the IPT$^*$ approximation has 
the same functional
form as in Eq.~(\ref{eq: sc_ipt}), but the bare Green's function $G_0$ is
replaced by the dressed Green's function $G$.

At half filling the Holstein model interaction is particle-hole symmetric,
so the Green's function and self energy are purely imaginary and the vertices
are real.  The self energy can be expressed by $\Sigma(i\omega_n)\equiv 
i\omega_nZ(i\omega_n)$, with $Z(i\omega_n)$ the renormalization function
for the self energy.  At half filling, both $\chi_m^0(X=-1)$ and
the maximal eigenvector of the
scattering matrix are also even functions of Matsubara frequency, so
the only contribution of the
irreducible vertex function to the eigenvalue of the scattering matrix
comes from the even Matsubara frequency
component $[\Gamma_{m,n}+\Gamma_{-m-1,n}]/2$.  The same result holds for the
SC channel at all fillings, because both the maximal eigenvector and the
bare susceptibility are symmetric functions of Matsubara frequency.

A comparison of these approximations to the exact QMC results for the electronic
self energy and one column of the irreducible vertex function in the CDW and
SC channels is made in Figures 3--5 for three different interaction
strengths at half filling.  The phonon frequency is set to be approximately
one-eighth of the effective bandwidth ($\Omega/t^*=0.5$) as was done in the
QMC solutions\cite{freericks_jarrell_scalapino}.  The energy cutoff is set to
include 256 positive Matsubara frequencies for the perturbative approximations.

At weak coupling $(g=0.4$, $T=0.0625$, 
Figure 3), the second-order IPT approximation
is a reasonable approximation to the electron self energy [Figure 3~(a)]
and to the CDW irreducible vertex function [Figure 3(b)].  Note that the IPT
underestimates the self energy (an effect that raises $T_c$) and overestimates
the magnitude of the vertex (an effect that also raises $T_c$), so one 
expects it to overestimate the CDW transition
temperature (both the IPT and the IPT$^*$ approximations yield similar results
for the irreducible vertex function).  
Clearly third-order diagrams play an important role even at this
weak a value of the coupling strength.  In this sense, the IPT is less
accurate for retarded interactions than it is for instantaneous interactions
(see Section IV).

As the coupling strength is increased to the point where a double-well 
structure begins to develop in the effective phonon potential of the QMC 
simulations of Ref.~\onlinecite{freericks_jarrell_scalapino} ($g=0.5$,
$T=0.111$) 
one can see strong-coupling effects begin to enter into the QMC vertex.  
The IPT still underestimates the self energy [Figure 4~(a)], but as the 
strong-coupling regime is approached, the vertices become increasingly 
attractive [Figure 4~(b)].  This enhanced attraction is not represented by 
either the IPT or the IPT$^*$ approximations.  Note that in this regime the 
self energy and the vertices are underestimated, which implies that the 
transition temperature may be determined more accurately than expected, because 
these effects tend to cancel each other out in determining $T_c$.  The 
underestimation of the magnitude of the vertex will eventually cause $T_c$ to 
drop due to the large self-energy renormalization.

When one is well into the strong-coupling regime ($g=0.625$, $T=0.167$,
Figure 5) the
deficiencies of the IPT approximation become more apparent.  The self energy
is underestimated by almost a factor of two [Figure 5~(a)], 
and the vertices are underestimated by almost a factor of three at intermediate
frequencies [Figure 5~(b)].  Even at this large a value of the
coupling, both the IPT and IPT$^*$ approximations remain similar to
each other.  At this point one would say the approximate theory is
failing to faithfully represent the exact solution.

A comparison of the IPT approximation to either a second-order conserving
approximation or to ME theory\cite{freericks_conserving} shows that the IPT is
a superior approximation for both the self energy and the vertices, but none
of these approximate methods is accurate over a wide range of interaction
strength, indicating once again the importance of the third-order diagrams.

At half filling, the Holstein model always has a transition to a CDW-ordered
phase at ${\bf q}=(\pi ,\pi ,\pi ...)$ $(X=-1)$.  The transition temperature
to this commensurate CDW is plotted in Figure 6 as a function of the
interaction strength.  Both second-order IPT approximations are compared
to the QMC simulations\cite{freericks_jarrell_scalapino}.  The IPT
and the IPT$^*$ approximations are very accurate in determining $T_c$; the
peak position and height are both reproduced well.  This result is
definitely a numerical coincidence, since the self energy and vertices are
poorly approximated at the peak of the curve $(g=0.625)$.  Furthermore, it
is the self energy renormalization that ultimately causes
$T_c$ to turnover, but $T_c$ drops 
far too rapidly in this strong-coupling regime.  This 
result is similar to what was found for a strong-coupling 
approximation\cite{freericks_strong}: there the transition temperature dropped
too rapidly in the weak-coupling regime.  

The IPT and IPT$^*$
approximations are superior to either a second-order
conserving approximation or to ME
theory in both a qualitative and a quantitative determination of the
CDW transition temperature at half filling.
They also have the correct limiting behavior as $U\rightarrow 0$
because they are second-order perturbation 
theories\cite{vandongen,martin_flores,freericks_conserving}.  

As the system is doped away from half-filling, the CDW instability remains
locked at the commensurate point $(X=-1)$ until it gives way
to a SC instability (incommensurate order may appear in a very narrow region
of phase space near the CDW-SC phase 
boundary\cite{freericks_jarrell_scalapino,feinberg} 
but is neglected here).  In Figure 7, the phase diagram of the 
Holstein model is plotted for three values of the interaction strength
($g=0.4$, $g=0.5$, and $g=0.625$).  The weak-coupling QMC 
data\cite{freericks_jarrell_scalapino} ($g=0.4$) are 
reproduced most 
accurately by the perturbative approximations, as expected from the
comparison of the self energy and the vertices in Figure 3. 
As the coupling increases to $g=0.5$ (where the double-well structure begins
to form in the effective phonon potential\cite{freericks_jarrell_scalapino}),
the approximations become less accurate.  The anomalous behavior of the 
CDW transition temperature not being a maximum at half filling already appears
for the IPT approximation, but the IPT approximation is more accurate at
determining the CDW-SC phase boundary.  This anomalous behavior in the CDW
transition temperature most likely occurs because the approximation is not
a conserving approximation.  Note that the SC transition temperatures are 
still accurately reproduced in this regime.

When one is well into the strong-coupling regime $(g=0.625)$ both the IPT
and IPT$^*$ approximations have large anomalous behavior near half filling.
The CDW-SC phase boundary is also poorly approximated.  Clearly both 
approximations are failing at this large a value of the coupling.

Both the IPT and IPT$^*$ approximations yield virtually identical results
for the Holstein model at a moderate phonon frequency.  The self energy
and the irreducible vertex functions are poorly reproduced, implying that
third-order diagrams are important when the interaction is retarded,
but remarkably, the transition temperatures are reproduced quite well.
At half filling both approximations show a peak in the CDW transition
temperature.  Off of half filling, the SC transition temperature is
determined more accurately than the CDW transition temperature (indeed
anomalous behavior enters into the CDW channel near half filling when
the interaction strength is large enough).  The phase boundary between
CDW and SC order is also accurately reproduced.

\section{Hubbard Model}

The Hubbard model in Eq. (\ref{eq: hubbham}) is the infinite-frequency
limit $(\Omega\rightarrow\infty)$ of the Holstein model.  The Hubbard model
has an electron-electron interaction that only occurs between electrons
with opposite spins.  This happens because of the cancellation of the
direct and exchange diagrams which causes all electron-electron interactions
between like-spin particles to vanish.  The perturbation theory becomes
much simpler in the Hubbard model case, because of this reduction of
diagrams, and can be performed to higher order.  Here the truncated
IPT approximation will be carried out to fourth order, and will be
compared to the QMC simulations\cite{jarrell} to determine their accuracy.  
Previous work has concentrated on second-order
conserving approximations\cite{muellerhartmann,hirashima}, 
third- and fourth-order conserving approximations\cite{freericks_conserving}, 
or the fluctuation-exchange (FLEX)
approximation\cite{muellerhartmann,freericks_conserving}.

One expects that a truncated approximation will be superior to an infinite
summation of random-phase approximation bubbles and particle-hole
and particle-particle ladders because the many-body problem reduces to 
a self-consistently embedded Anderson impurity model, and the analysis
of Yosida and Yamada\cite{yosida_yamada} has shown that the total 
fourth-order corrections to the
self energy are an order of magnitude smaller and opposite in sign to the
fourth-order contribution of the FLEX approximation.  The irreducible
vertex functions should have similar effects, but have not yet been analyzed
in detail.

%%JIM I rearranged the order of this paragraph some
The diagrammatic expansion for self energy (in the IPT approximation)
of the Hubbard model is given in Figure 8.  Here, the solid lines 
denote the bare Green's function $G_0$ and the dotted lines denote the
instantaneous Coulomb interaction $U$.  The first line includes the 
first-order Hartree contribution (which is a constant that is absorbed
into a renormalized chemical potential), the second-order contribution
and the third-order particle-hole and particle-particle ladders.  The
second line contains the fourth-order contributions from the RPA bubbles and
the particle-hole and particle-particle ladders.  The third and fourth
lines include all of the remaining diagrams that enter at
fourth-order.  The fifth line contains the insertion of the second-order
self energy into the second-order diagrams.   The irreducible vertex 
functions are too cumbersome to represent diagrammatically.
Explicit formulas for the electronic self energy
of the Hubbard model thru fourth order and for the CDW vertex have been given 
before\cite{freericks_conserving} (except for the second-order self-energy
insertions into the second-order diagrams which are easily determined).
Once again IPT denotes the approximation where the irreducible vertex 
is is strictly truncated at a given order, while the IPT$^*$ approximation
determines the vertex functions by using the dressed Green's functions.

Since the Hubbard-model interaction is particle-hole symmetric, the half-filled
band corresponds to $\mu=0$, and the Green's functions are purely imaginary.
The odd-order contributions to the self energy all vanish 
and each of the fourth-order contributions on a given line in Figure 8 are
identical\cite{yosida_yamada}.
Since the self-energy is an even function of $U$ at half-filling,
but the irreducible vertex function contains both even and odd powers of
$U$, the only difference between a truncated approximation of
order $2n$ and of order $2n+1$ is that the irreducible vertex function
is larger for the odd-order approximation.  Therefore, we expect
that an even order approximation will underestimate the transition temperature
(in weak-coupling) and an odd-order approximation will overestimate $T_c$.

One of the most remarkable results of Yosida and Yamada is that the total
fourth-order contribution to the self energy is an order of magnitude smaller
than any of its component pieces.  One can ask if the same near cancellation
holds in the mapping to the infinite-dimensional lattice.  The answer is
affirmative, as can be seen in Figure 9, where each of the individual
contributions to the self energy are plotted at half-filling for $U=-1$ and
$T=0.05$.  The dashed line shows the second-order contribution to the
self energy.  The dotted lines are each of the four different fourth-order
contributions.  FLEX denotes the first fourth-order contribution corresponding
to the fourth-order bubbles and ladders. Line 2 and line 3 denote the 
contributions arising from the next two lines respectively
in Figure 8, and IPT is used
to denote the contributions arising from the second-order self-energy
insertions into the second-order diagram (the last line of Figure 8).  The 
total of all of the fourth-order
diagrams is plotted with the solid line.  One can see that the remarkable
near cancellation of the fourth-order diagrams still holds in 
infinite-dimensions, and in this case the fourth-order contribution is the
same sign, but more than an order of magnitude smaller than the FLEX
contribution at fourth order.

A comparison of the different approximation schemes versus the exact QMC
results are given in Figures 10--12 for three
different values of $U$.  The second-order and third-order approximations
employ an energy cutoff of 256 positive Matsubara frequencies; the 
fourth-order approximation uses 64 positive Matsubara frequencies.
In Figure 10~(a) the self-energy renormalization
function is plotted for the two different approximations at $U=-1$ ($T=0.05$).
Note that the fourth-order approximation virtually reproduces the QMC
results.  In Figure
10~(b) the even component of one row of the irreducible vertex function
for the CDW channel are compared for $U=-1$ in the IPT approximation.
All of the approximations are in reasonable agreement with the QMC results,
but the accuracy is significantly reduced relative to what was found for
the self energy.  The IPT$^*$ approximation is compared to the QMC results in
Figure 10~(c).  These results are similar to the IPT approximation, but
perhaps not as accurate.

As the coupling strength is increased to $U=-2$ ($T=0.125$)
the self energy is becoming 
bracketed by the two low-order approximations, with the fourth-order
approximation overestimating the self energy, and the second-order approximation
underestimating the self energy, as shown in Figure 11~(a).  The CDW vertex is 
plotted in Figure 11~(b) for the IPT approximation.  One can see strong-coupling
effects beginning to enter as the QMC vertex becomes more attractive at
low frequency transfer than predicted by the even-order approximations.
The third-order IPT approximation is producing the correct qualitative shape
of the vertex, but it is overestimating the magnitude by a large amount.
The IPT$^*$ approximation is shown in Figure 11~(c).  Here 
the two different methods for calculating the vertex yield quite
different results.  In the IPT$^*$ approximation, all orders are showing
a weakening of the vertex at lower frequencies, which is not present in
the QMC data.

Finally, at a rather strong value of the coupling $(U=-3$, $T=0.167)$, where the
transition temperature reaches its maximal value, the self energy is still
reproduced quite accurately by both orders of the IPT [see Figure 12~(a)].
The vertices are, however, quite poorly approximated.  In Figure 12~(b) and
12~(c) the results for the IPT and IPT$^*$ respectively are presented.  The 
QMC vertex is becoming strongly attractive here.  Only the third-order
IPT approximation has the correct qualitative behavior, but it overestimates
the magnitude of the vertex.  The IPT$^*$ approximation produces a fairly
flat frequency dependence to the vertex which is quite accurate at all but
the lowest frequency transfers (where it is almost an order of
magnitude too small).

The conclusion that we can draw from this comparison of the self energy and
irreducible vertex functions is that the IPT approximation is quite
accurate for the self energy, but is a rather poor approximation for the
irreducible vertex function.  It is known that the reason why the IPT is so
accurate for the self energy arises from the fact that it properly reproduces
the atomic limit\cite{ferrer_martinrodero_flores,kotliar_hubb}, but it
is possible that it does not produce the atomic limit for the
irreducible vertex functions.  If this were the case, one could try another
{\it ad hoc} interpolation scheme that interpolates between the perturbative
result for the vertex, and the atomic limit, to provide a more 
accurate approximation.  

Note also that since the self energy is reproduced
quite accurately by the IPT, but the magnitude of the vertex is grossly
underestimated, the transition temperature will drop too rapidly in
the strong-coupling regime (with the exception of the third-order IPT 
approximation where the transition temperature will grow too rapidly).
This is illustrated by plotting the transition
temperature for the CDW transition at half-filling in the
attractive Hubbard model as a function of $U$ (Figure 13) and comparing to
the QMC solution\cite{jarrell}.
In the IPT approximation [Fig.13~(a)], the peak in $T_c$ as a function
of $U$ is properly reproduced for the even-order approximations, but not for the
odd-order one.  The transition temperature decreases too rapidly in the
strong-coupling regime because of the underestimation of the vertex.  
In the IPT$^*$
approximation [Fig. 13~(b)], all orders have a peak in $T_c$ as a function 
of $U$.  The
third-order approximation is the most accurate here, because it has the most
attractive vertex, but even in this case the quantitative
agreement with the QMC is not
as good as was found for the truncated conserving 
approximations\cite{freericks_conserving}, where the transition temperature 
did not have a peak, but the underestimation of the self-energy compensated for
the underestimation of the vertex to produce reasonably accurate transition
temperatures into the strong-coupling regime.

The doping dependence of the SC transition temperature is plotted in Figure 14 
for both the second and third order IPT and IPT$^*$ approximations and $U=-1$.
Even at this weak a value of the coupling strength, the second-order IPT and
IPT$^*$ approximations show the anomalous behavior of an increase in $T_c$
for small dopings off of half filling.  A third-order approximation works
much better (because third-order contributions to the self energy enter
when the electron concentration is not equal to 1), and does not display
the anomalous behavior (at this value of $U$).  Once again, the IPT$^*$
approximation is more robust against the anomaly in $T_c$, but all IPT
approximations quickly fail as the coupling strength increases beyond
$U=-1$ (because they fail at half filling as shown in Fig. 13).

\section{Conclusions}

The iterated perturbation theory approximation\cite{georges_kotliar}
has been applied to the attractive Holstein and Hubbard models in infinite
dimensions.  This weak coupling theory is not a conserving approximation
because it corresponds to a strictly truncated perturbation series for the
self energy.  The vertex functions are either approximated with a strictly
truncated perturbation series too (IPT), or with a truncated perturbation
series that includes all self energy insertions (IPT$^*$).

At half filling, these approximate theories are accurate in reproducing 
the electronic self energy (for the Holstein model, the accuracy increases as 
the phonon frequency increases) but do not reproduce the correct behavior of 
the irreducible vertex functions in the strong-coupling regime.  

In particular, for the half-filled Hubbard model, the IPT is a very accurate 
approximation for the self energy of the Hubbard model, but is a poor 
approximation for the irreducible vertex functions a low frequency transfer.  
As a result, the transition temperature does in general display a peak 
(because of the large self-energy renormalizations), but decreases too rapidly 
in the strong-coupling regime.

The IPT is a much less accurate approximation for the self energy of the
Holstein model at finite phonon frequency (indicating the importance 
of higher-order diagrams).  The IPT is also a poor approximation for the
vertex functions, but, remarkably, the transition temperatures are quite
accurately approximated both at half filling and off of half filling (for
small enough coupling strength).  The peak position and height of $T_c$ at
half filling are reproduced well, as is the doping dependence of $T_c$
up to the region of parameter space where a double-well structure begins to 
develop in the effective phonon potential.

As both of these systems are doped away from half filling, the transition 
temperatures (to either a CDW or a SC) display the anomalous behavior of 
initially {\it increasing} with doping.  This anomalous behavior first enters 
at some critical value of the coupling, and becomes worse as the coupling 
strength is increased. For the Holstein model (with $\Omega=0.5$) this
%% JIM note the changes to \approx instead of = and <
critical coupling lies near $g\approx 0.5$, whereas for the Hubbard model it
occurs at $|U|\approx 1$.  Third-order diagrams tend to push this anomaly out
to larger values of interaction strength as expected because the third-order
contributions to the self energy can be large as the system is doped off
of half filling.

As a final summary, the IPT is a highly accurate approximation for the
self energy when the interaction is instantaneous, but generally underestimates
the magnitude of the vertex functions at low frequency transfer; it does not
approximate the transition temperature well in the strong-coupling regime.
The IPT is a less accurate approximation for both the self energy and vertices
for retarded interactions, but is much more accurate in determining transition
temperatures because the errors in the self energy and the vertices
tend to cancel in the determination of $T_c$.  
The the explanation for this behavior is that the IPT is
an accurate approximation when third-order terms can be neglected.  This
occurs for the self energy of the Hubbard model at half filling, but is
not true for the self energy when the interaction is retarded or the filling
is different from 1, or for the irreducible vertex functions.
Qualitatively the IPT approximation is better than the same
order conserving approximation\cite{freericks_conserving} for the self energy
of the Hubbard model, but the transition temperatures are better approximated
(quantitatively) by the conserving approximation.  For the Holstein model, the
IPT is superior to the conserving approximation in all aspects except for the 
anomalous behavior in $T_c$ near half filling and for large enough coupling
strength.

\acknowledgments

We would like to thank N. E. Bickers 
for many useful discussions.  We would especially like to thank D. Scalapino
for his continued interest in this problem and for numerous discussions.
This research is supported by the Office of Naval Research under Grant No.
N00014-93-1-0495, the NSF grant No. DMR DMR-9107563, and by the Ohio 
Supercomputing Center.  In addition, MJ would like to thank the NSF
NYI program for their support.

%\end{document}

\begin{figure}
  \caption{Dyson equation for the self energy of the Holstein model.
The solid lines denote the {\it bare} (electronic)
Green's function and the wavy lines denote the phonon propagator.
The self energy includes the Hartree and Fock contributions, 
the second-order dressing of the phonon line, the lowest-order vertex
correction, and the second-order rainbow diagram (which arises
from the self-energy insertion of the Fock diagram into the Fock diagram). }
  \label{fig:1}
\end{figure}

\begin{figure}
\caption{The irreducible vertex functions in the CDW and SC channels.
The CDW irreducible vertex function is shown in (a).  
Note that the vertex corrections
(exchange diagrams) modify the interaction {\it to lowest order} in the
CDW channel.  The SC irreducible vertex function is shown in (b).  
The vertex corrections first enter at second order in the SC channel.
In the IPT approximation the electron propagator is the bare Green's function 
$G_0$, while in the IPT$^*$ approximation the electron propagator is the
dressed Green's function $G$.}
\label{fig:2}
\end{figure}

\begin{figure}
\caption{ Comparison of the IPT (solid line) and IPT$^*$ (dashed line) to
the QMC solutions (solid dots)
for the Holstein model at half filling with phonon
frequency $\Omega=0.5t^*$, interaction strength $g=0.4t^*$, and temperature
$T=t^*/16$.  This example is generic for the weak-coupling limit.  In (a)
the self energy renormalization function $Z(i\omega_n)-1$ is plotted
against the Matsubara frequency.  In (b) the symmetric combination of the first 
column of the irreducible vertex function in the CDW channel
is shown.  Note that even at this weak a value of coupling strength the
third-order diagrams must play an important role.}
\label{fig:3}
\end{figure}

\begin{figure}
\caption{Comparison of the IPT (solid line) and IPT$^*$ (dashed line) to
the QMC solutions (solid dots)
for the Holstein model at half filling with phonon
frequency $\Omega=0.5t^*$, interaction strength $g=0.5t^*$, and temperature
$T=t^*/9$.  This example is generic for the transition region to the
strong-coupling limit.  The self-energy renormalization function (a) and
the irreducible vertex function in the CDW channel (b)
are pictured.  Note that in the limit
where the strong-coupling effects begin to manifest themselves, the QMC vertex 
becomes increasingly attractive at low frequency transfer.}
\label{fig:4}
\end{figure}

\begin{figure}
\caption{Comparison of the IPT (solid line) and IPT$^*$ (dashed line) to
the QMC solutions (solid dots)
for the Holstein model at half filling with phonon
frequency $\Omega=0.5t^*$, interaction strength $g=0.625t^*$, and temperature
$T=t^*/6$.  This example is generic for the
strong-coupling limit.  The self-energy renormalization function (a) and
the irreducible vertex function in the CDW channel (b)
are pictured.  Note that in the strong-coupling regime the QMC vertex 
becomes quite attractive at low frequency transfer.}
\label{fig:5}
\end{figure}

\begin{figure}
\caption{Transition temperature to the CDW-ordered state at half filling in
the Holstein model at an intermediate phonon frequency $(\Omega=0.5t^*)$.
The IPT approximation (solid line) is compared to the IPT$^*$
approximation (dashed line) and the QMC results (solid dots).  Both 
approximations are similar to each other and are accurate at predicting the
peak height and position.  This occurs because the underestimation of
the self energy is cancelling the underestimation of the vertex in the
calculation of $T_c$.}
\label{fig:6}
\end{figure}

\begin{figure}
\caption{Phase diagram of the Holstein model with $\Omega=0.5t^*$ at three
different coupling strengths $(g=0.4,0.5,0.625)$.  The solid dots are the QMC
solutions with CDW order, and the open triangles are the QMC results with
SC order (the dotted lines are a guide to the eye). 
The kinks in the solid (IPT) and dashed (IPT$^*$) lines 
occur at the CDW-SC phase boundaries.  Note the anomalous behavior of the 
CDW transition temperature not being maximal at half filling sets into the
approximate theories as the coupling strength increases.}
\label{fig:7}
\end{figure}

\begin{figure}
\caption{Self energy diagrams for the Hubbard model thru fourth order.  The
solid lines denote the {\it bare} electronic Green's functions,
and the dotted lines are the Coulomb interaction.  The first two lines
contain all of the bubbles and ladders thru fourth order.  The middle
two lines are the remaining diagrams that enter first at fourth-order.  The
last line includes the second-order self-energy insertions into the second-order
diagrams.  At half filling the
odd-order contributions to $\Sigma$ vanish, and each of the three
fourth-order contributions on the same line yield the same contribution
to $\Sigma$.}
\label{fig:8}
\end{figure}

\begin{figure}
\caption{Near cancellation of the fourth-order contributions to the
self energy.  The second-order contribution to the self energy (dashed line)
is compared to each of the four fourth order contributions (dotted lines)
and the net fourth-order contribution (solid line) at $U=-t^*$, $T=t^*/20$, and 
half filling.  The dotted line denoted FLEX corresponds to the first line
of the fourth-order contributions in Figure 8.  The labels line 2 and line 3 
denote the next two lines of fourth-order contributions in Figure 8, and
the label IPT refers to the last line in Figure 8 corresponding
to the second-order self-energy insertions into the second-order diagrams.
Note that the total contribution of the fourth order diagrams is an order
of magnitude smaller than any of the individual classes of diagrams in Figure 8,
and that it is the same sign as the FLEX contributions.}
\label{fig:9}
\end{figure}

\begin{figure}
\caption{Comparison of the IPT and IPT$^*$ approximations to the QMC
solutions for the Hubbard model at half filling in the limit of weak-coupling
($U=-t^*,T=t^*/20)$.  The second-order (dashed line), third-order (dotted line),
and fourth-order (solid line) approximations
are compared the QMC solutions
(solid dots).  In (a) the self-energy renormalization function is plotted
against Matsubara frequency.  In (b) the even component of the first column
of the irreducible vertex function in the CDW channel is plotted in the
IPT approximation, and in (c) it is plotted in the IPT$^*$ approximation.}
\label{fig:10}
\end{figure}

\begin{figure}
\caption{Same as in Figure 10, but with a stronger value of the coupling
$(U=-2t^*,T=t^*/8)$.  The self energy is still reproduced accurately
by all orders, but only the third-order IPT approximation has the correct
qualitative behavior for the vertex.}
\label{fig:11}
\end{figure}

\begin{figure}
\caption{Same as in Figure 10, but with a stronger value of the coupling
$(U=-3t^*,T=t^*/6)$.  The self energy is still reproduced accurately
by all orders, but only the third-order IPT approximation has the correct
qualitative behavior for the vertex.}
\label{fig:12}
\end{figure}

\begin{figure}
\caption{Transition temperature to the CDW-ordered (and SC-ordered)
state in the Hubbard
model at half filling.  The second-order (solid line), third-order (dotted
line), and fourth-order (solid line) IPT (a) and IPT$^*$ (b) approximations are 
compared to the  QMC results (solid dots). Note that the transition temperature
curves all have a turnover (except for the third-order IPT approximation)
because the vertex is underestimated as the coupling strength increases.
The third-order IPT approximation does not turnover because the vertex is
overestimated. }
\label{fig:13}
\end{figure}

\begin{figure}
\caption{Transition temperature to the SC state in the Hubbard model as 
a function of electron concentration at $U=-1$.  The second-order IPT
(solid line), third-order IPT (dotted line), second-order IPT$^*$ (dashed
line), and third-order IPT$^*$ (chain-dotted line) are compared to
the QMC results (open triangle). Note that both second-order approximations 
display anomalous behavior near half filling even for this small a value of
the interaction strength.}
\label{fig:14}
\end{figure}

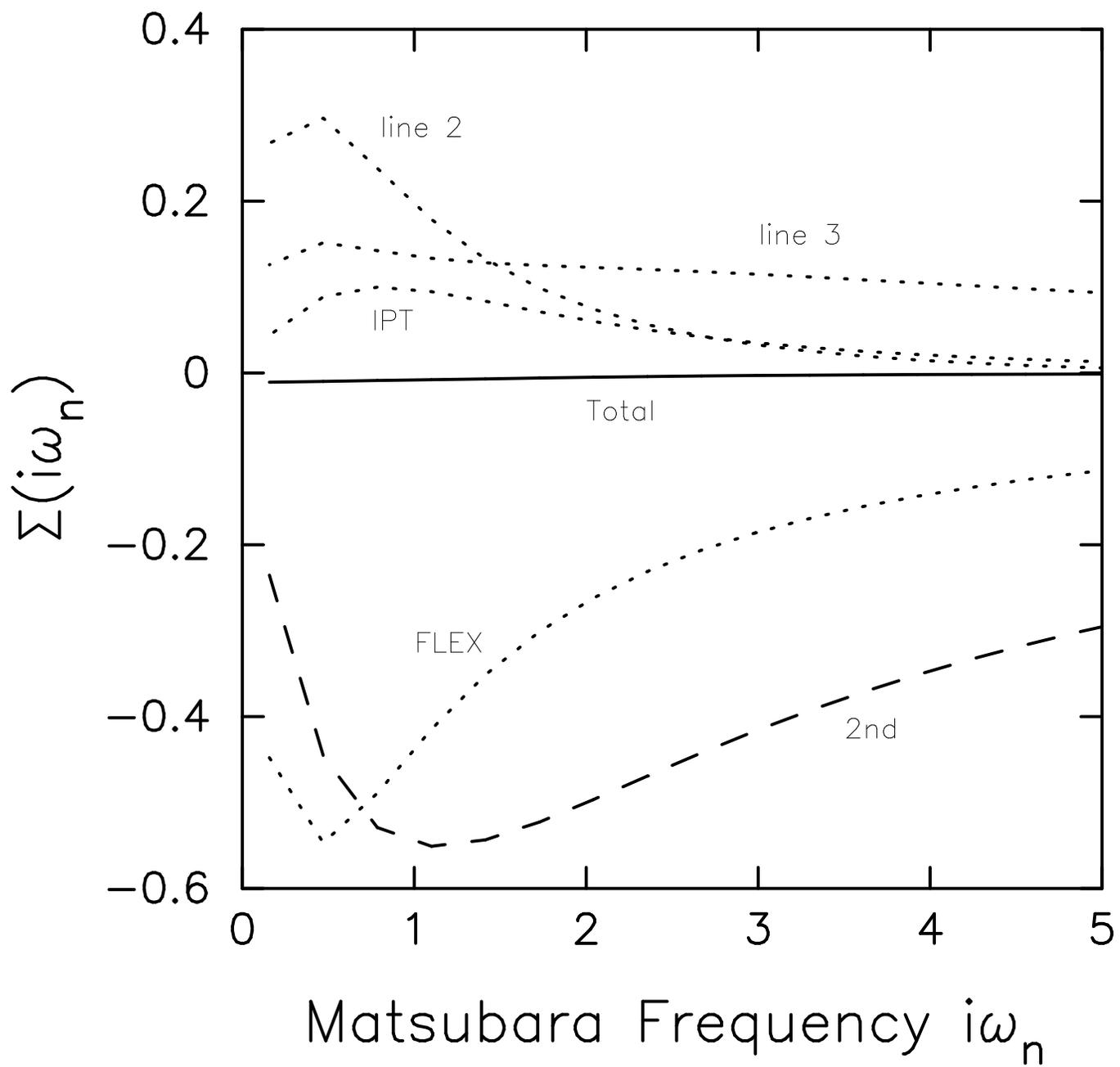
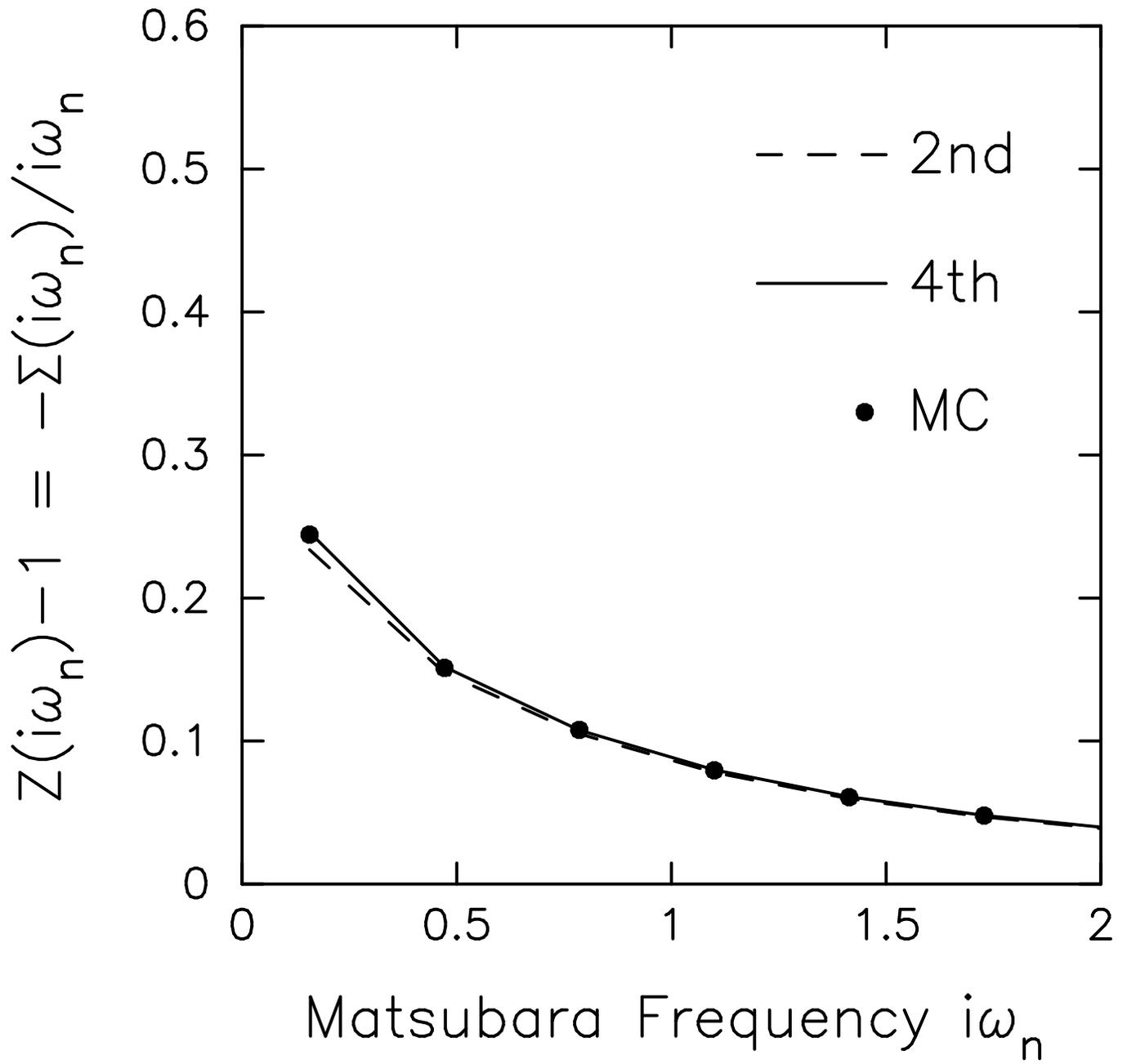
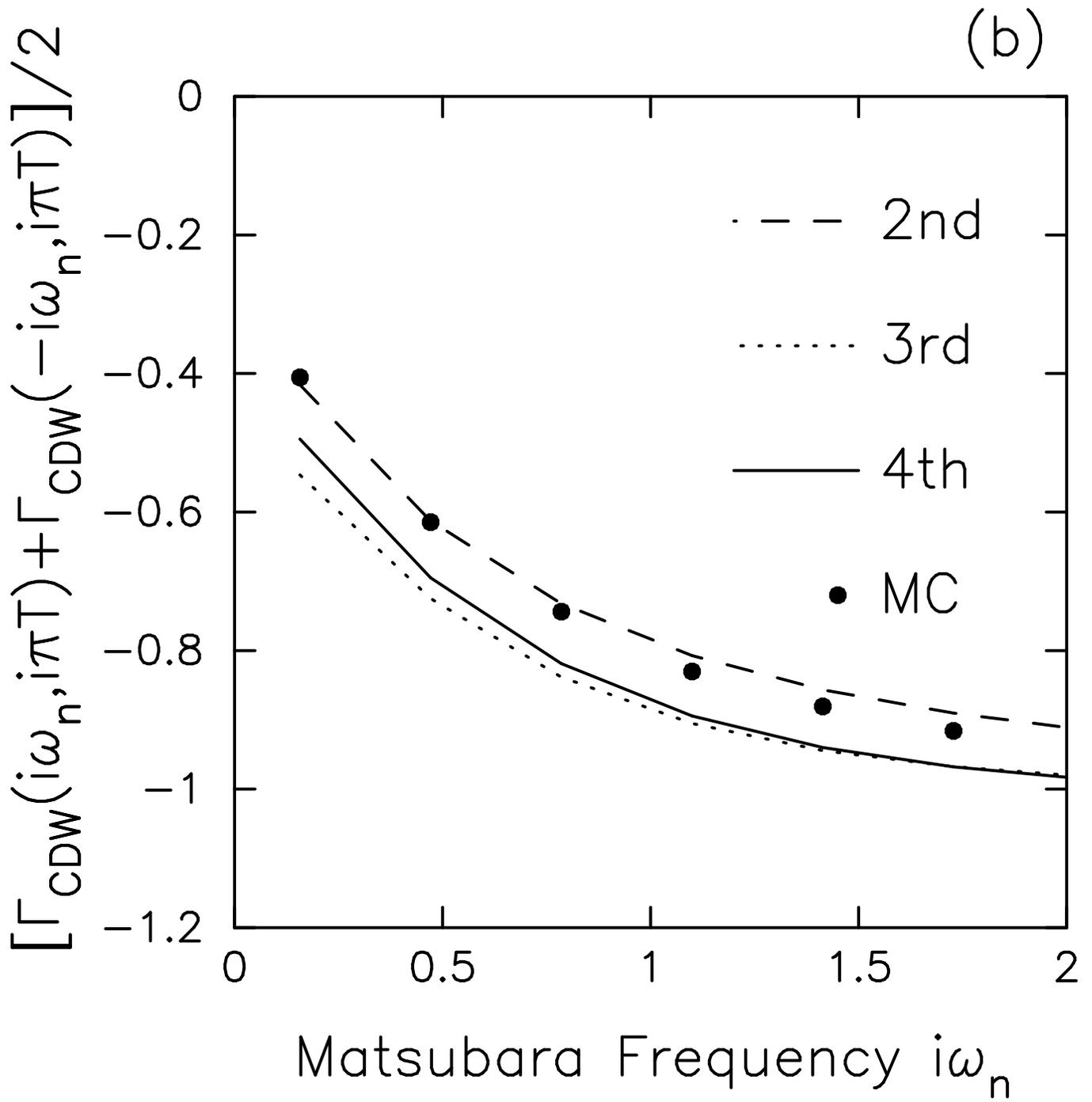
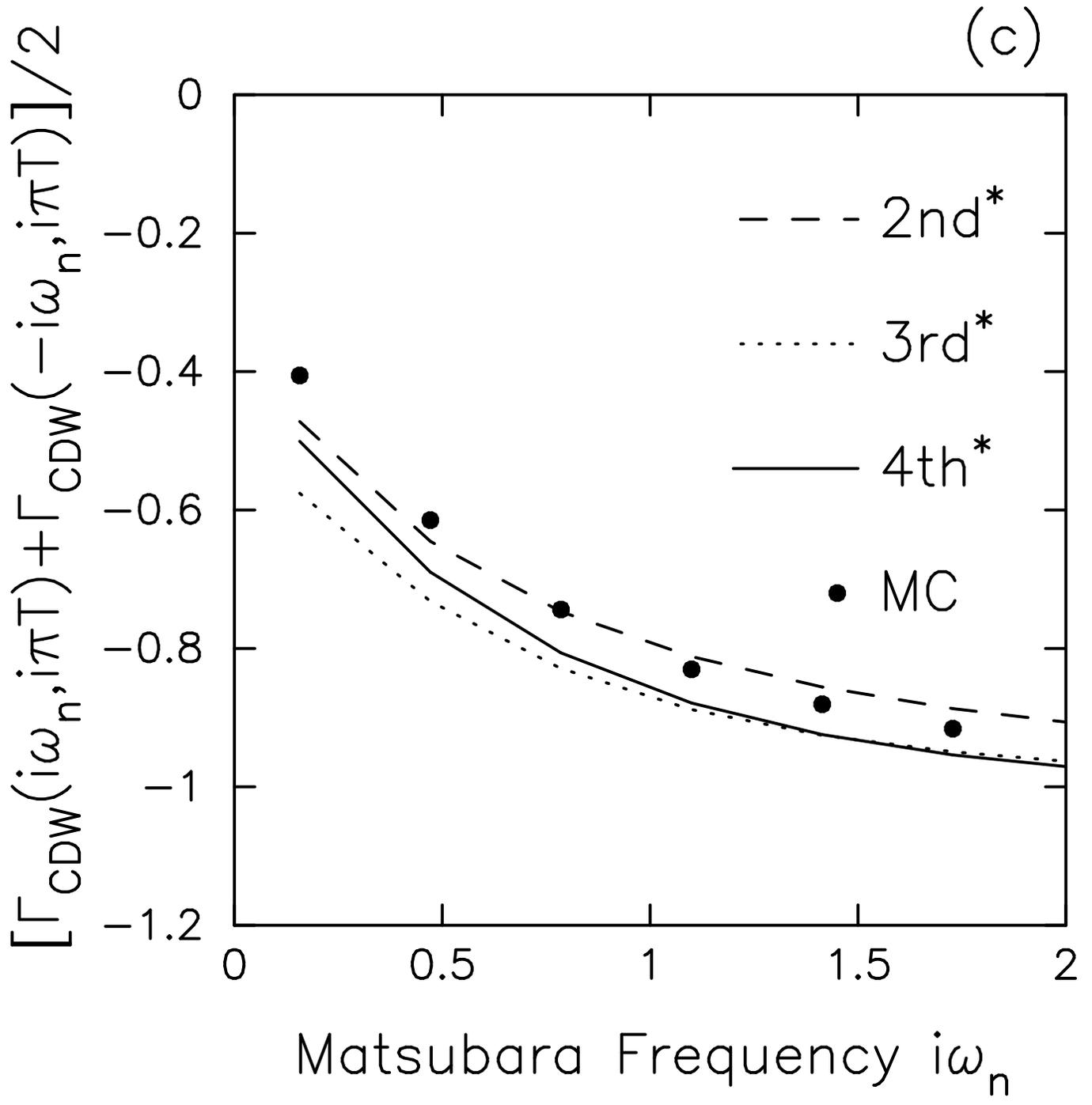
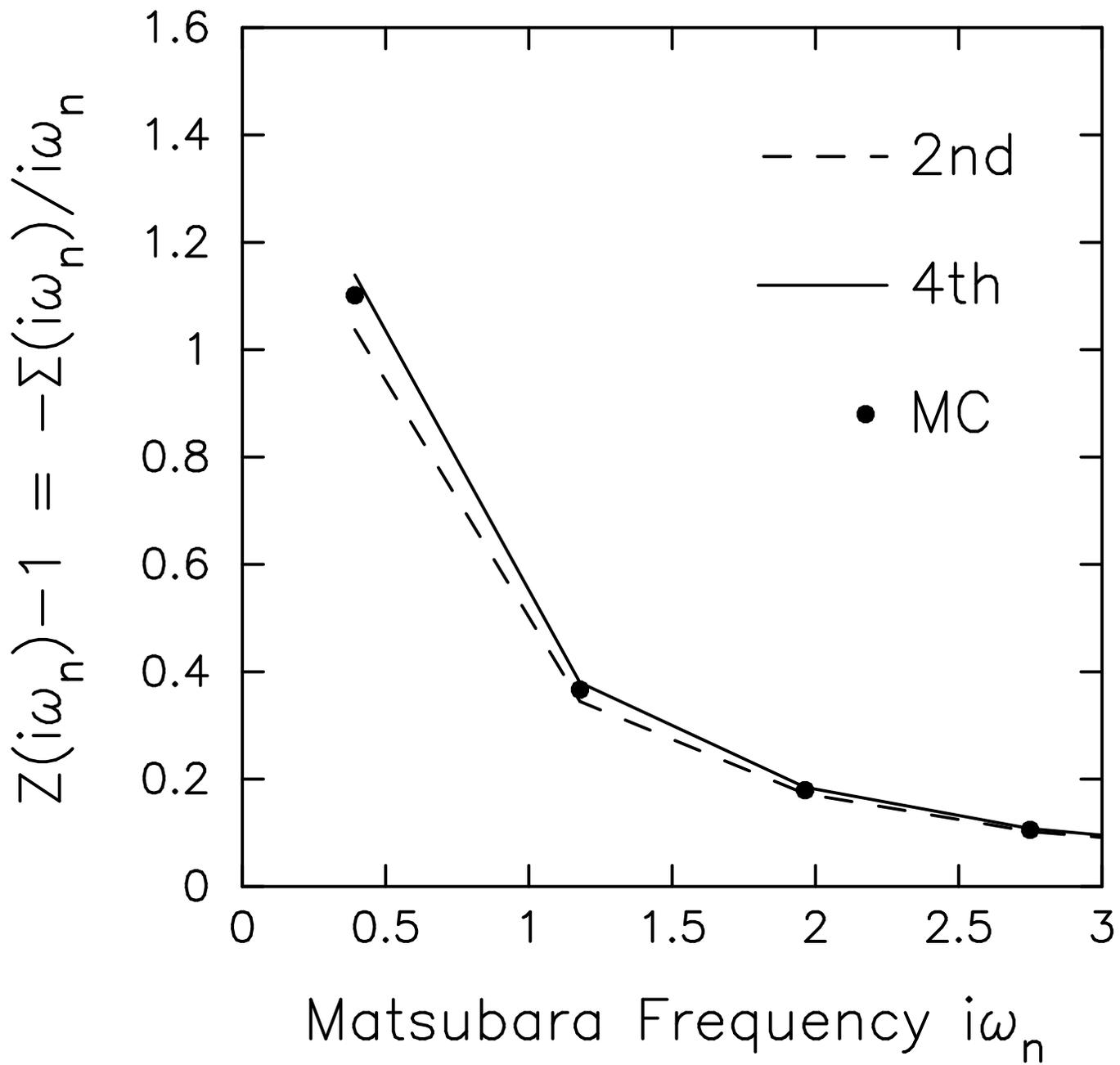
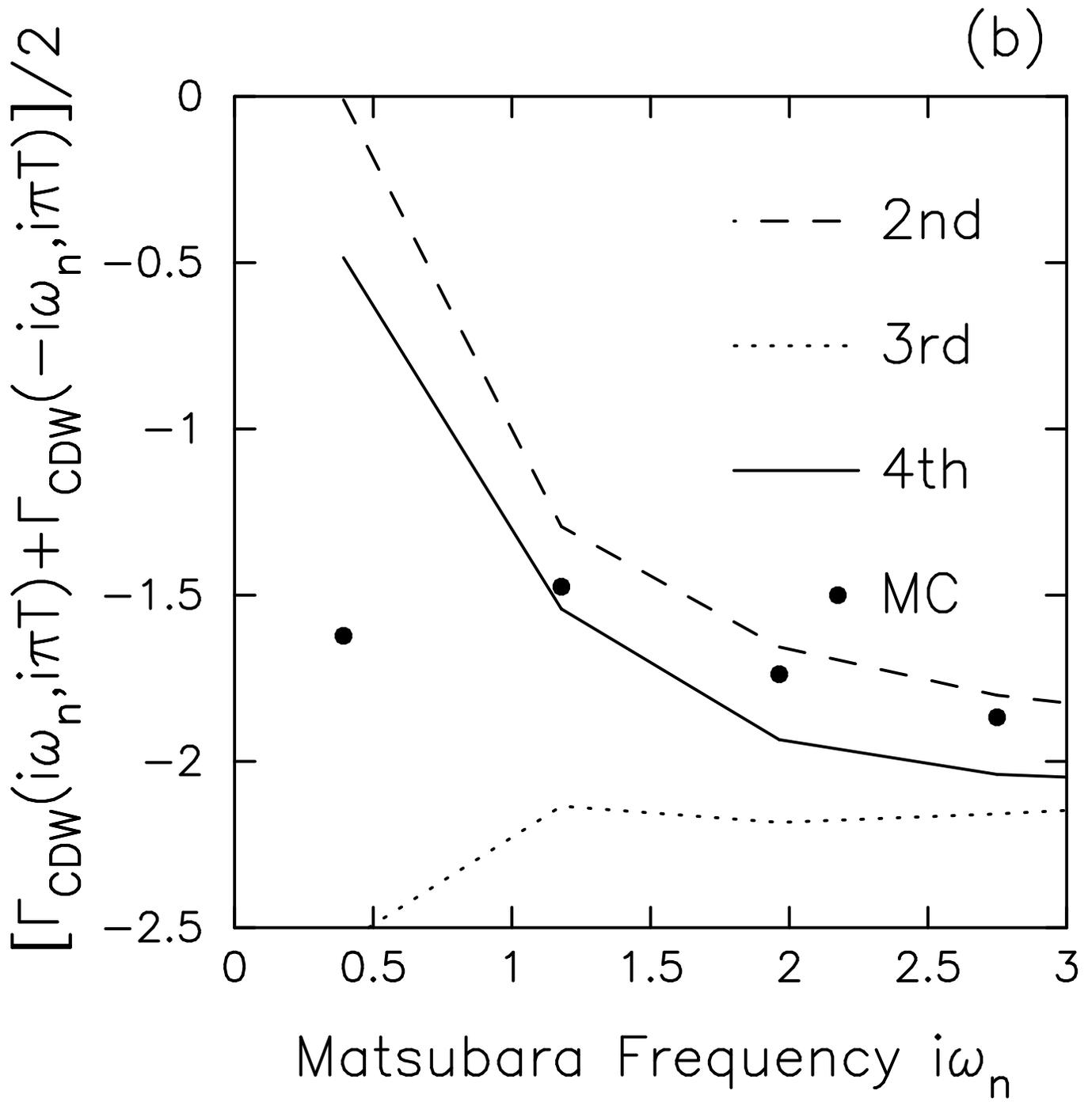
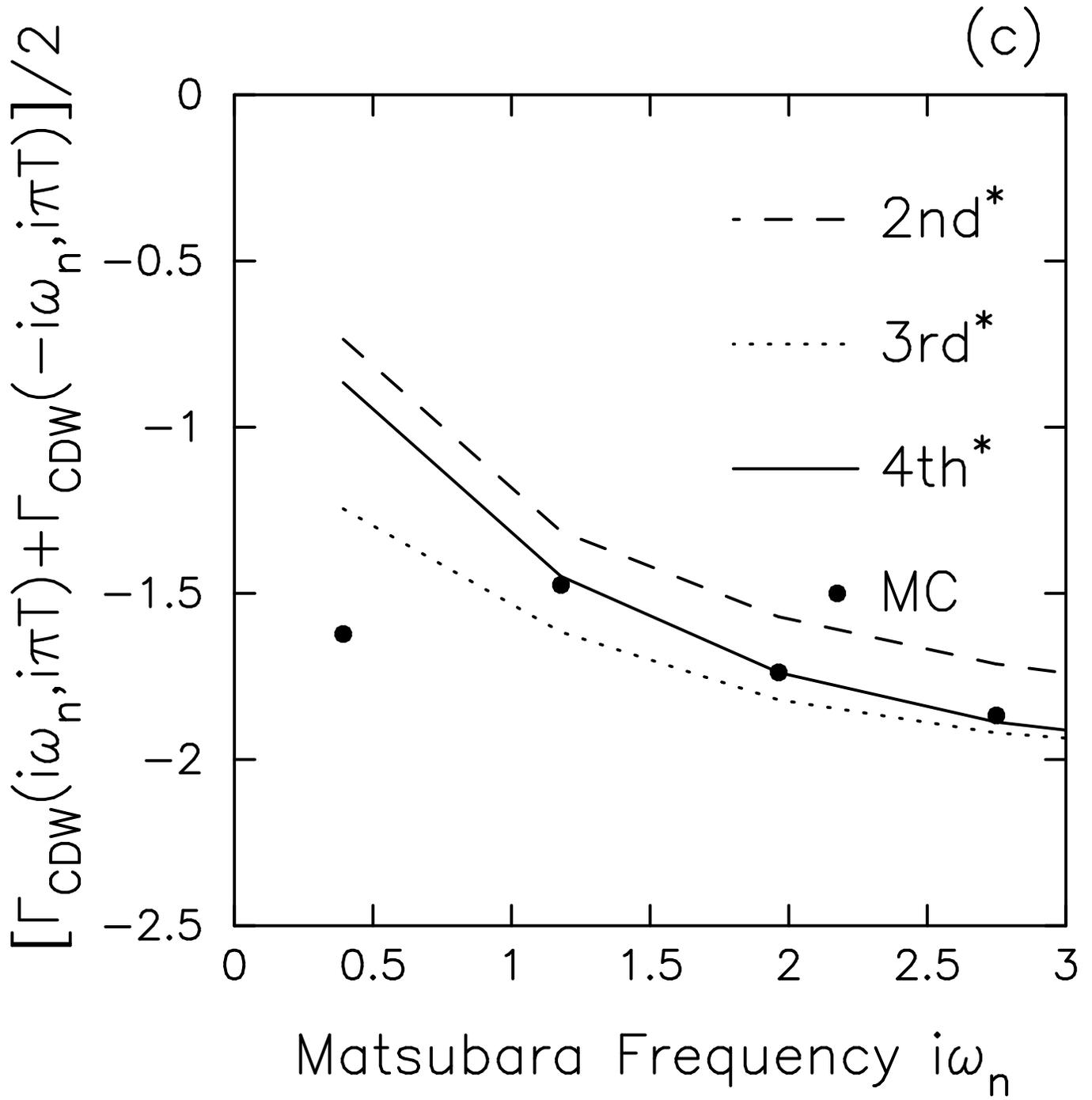
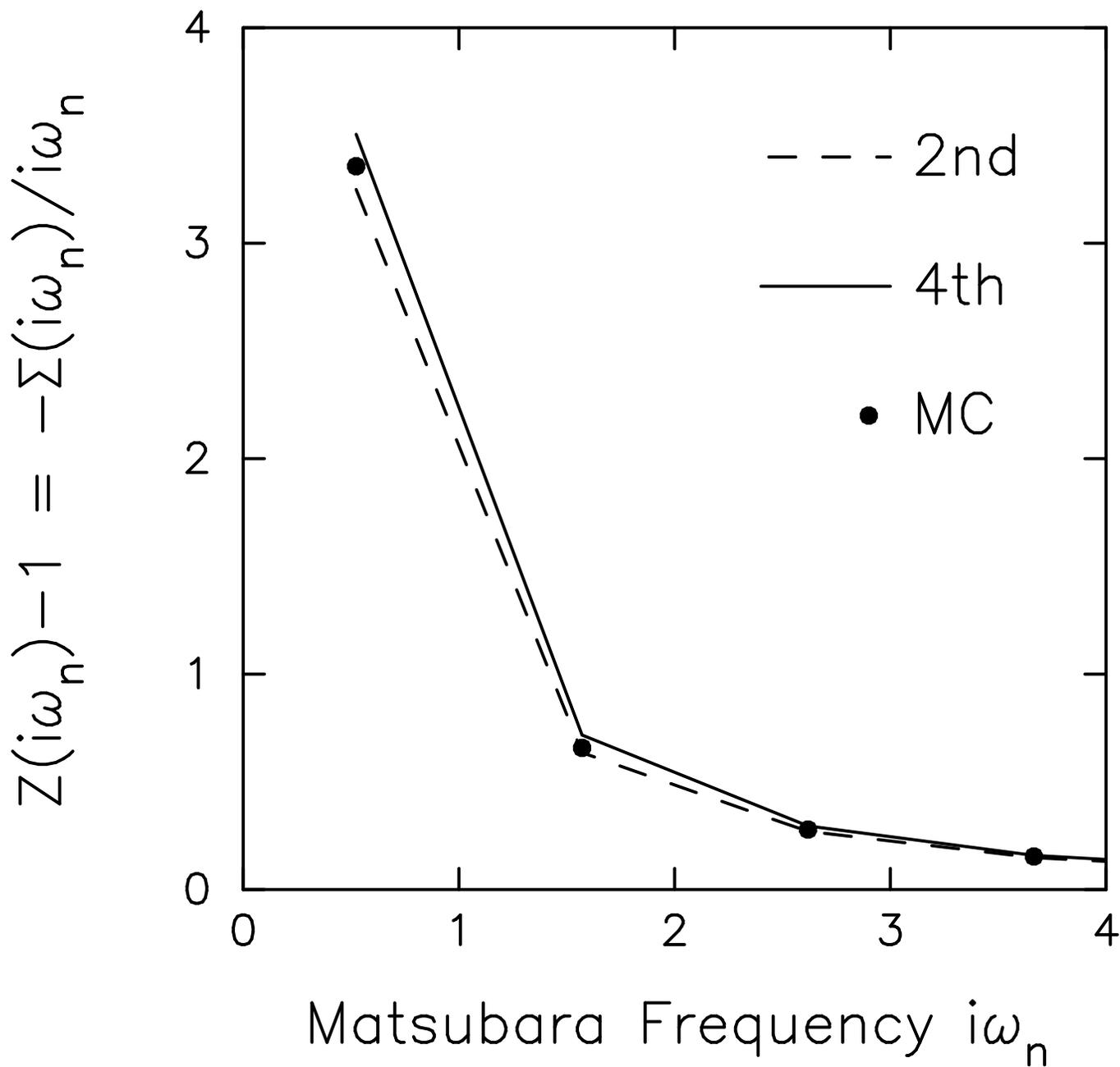
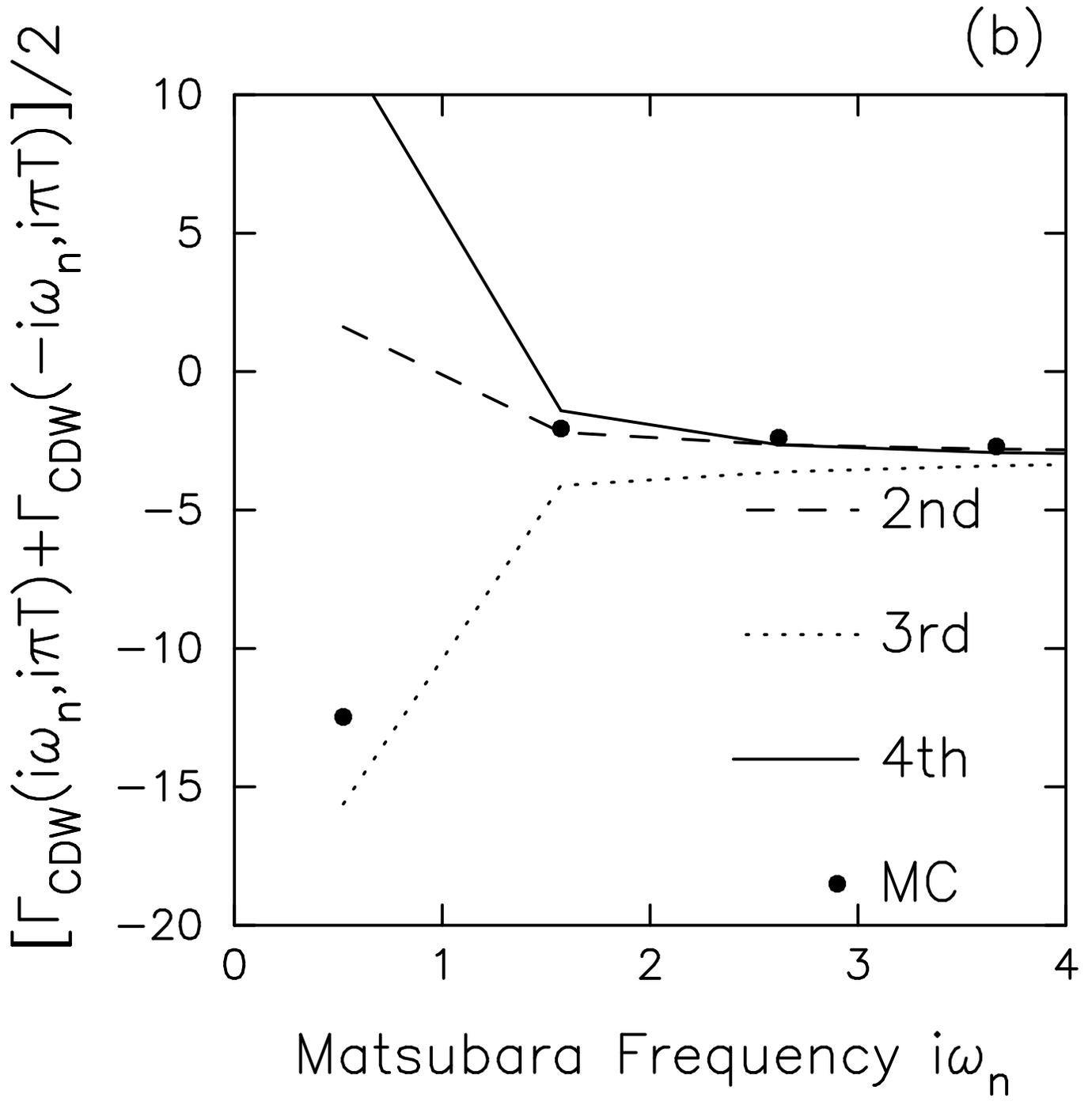
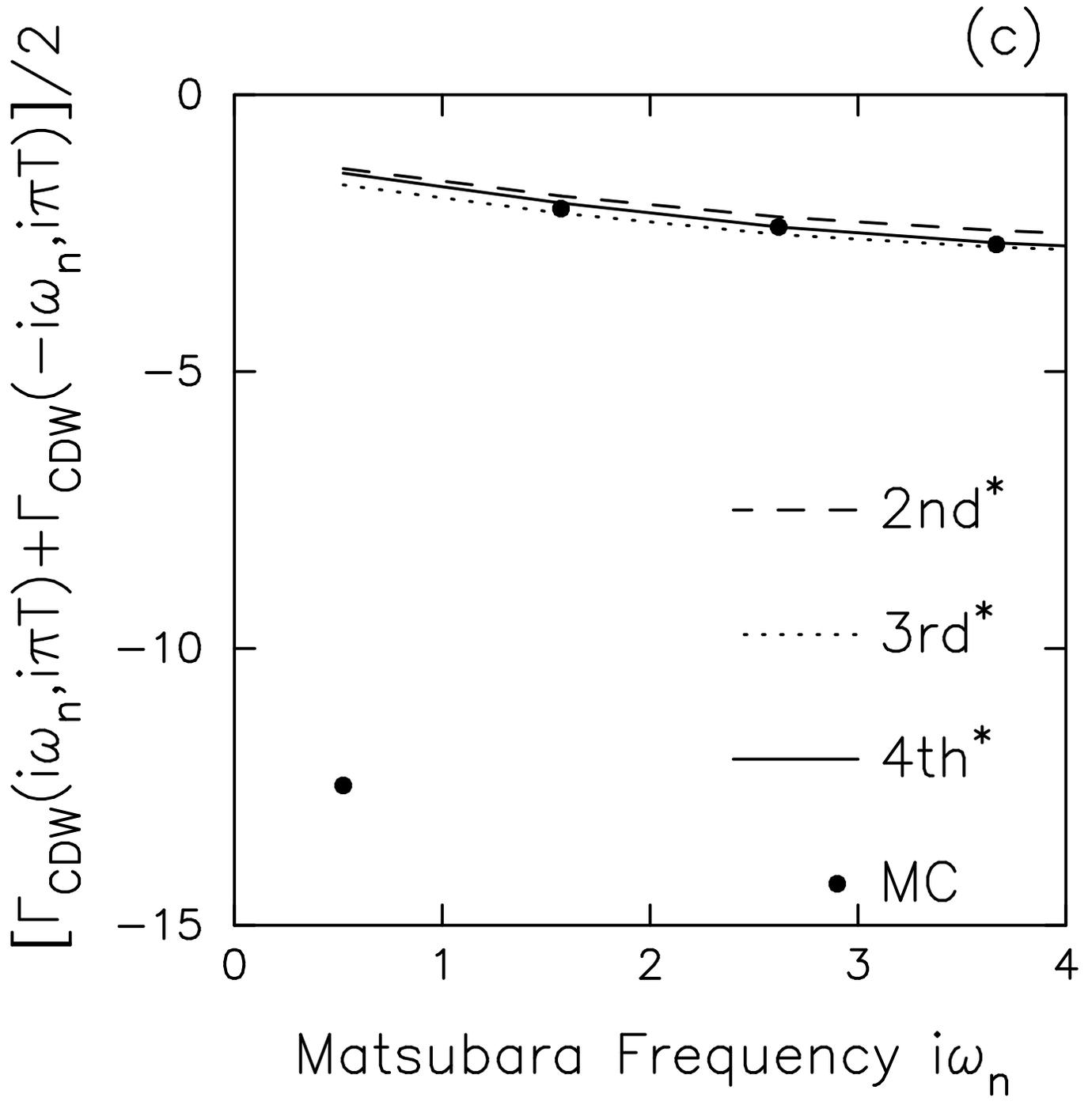
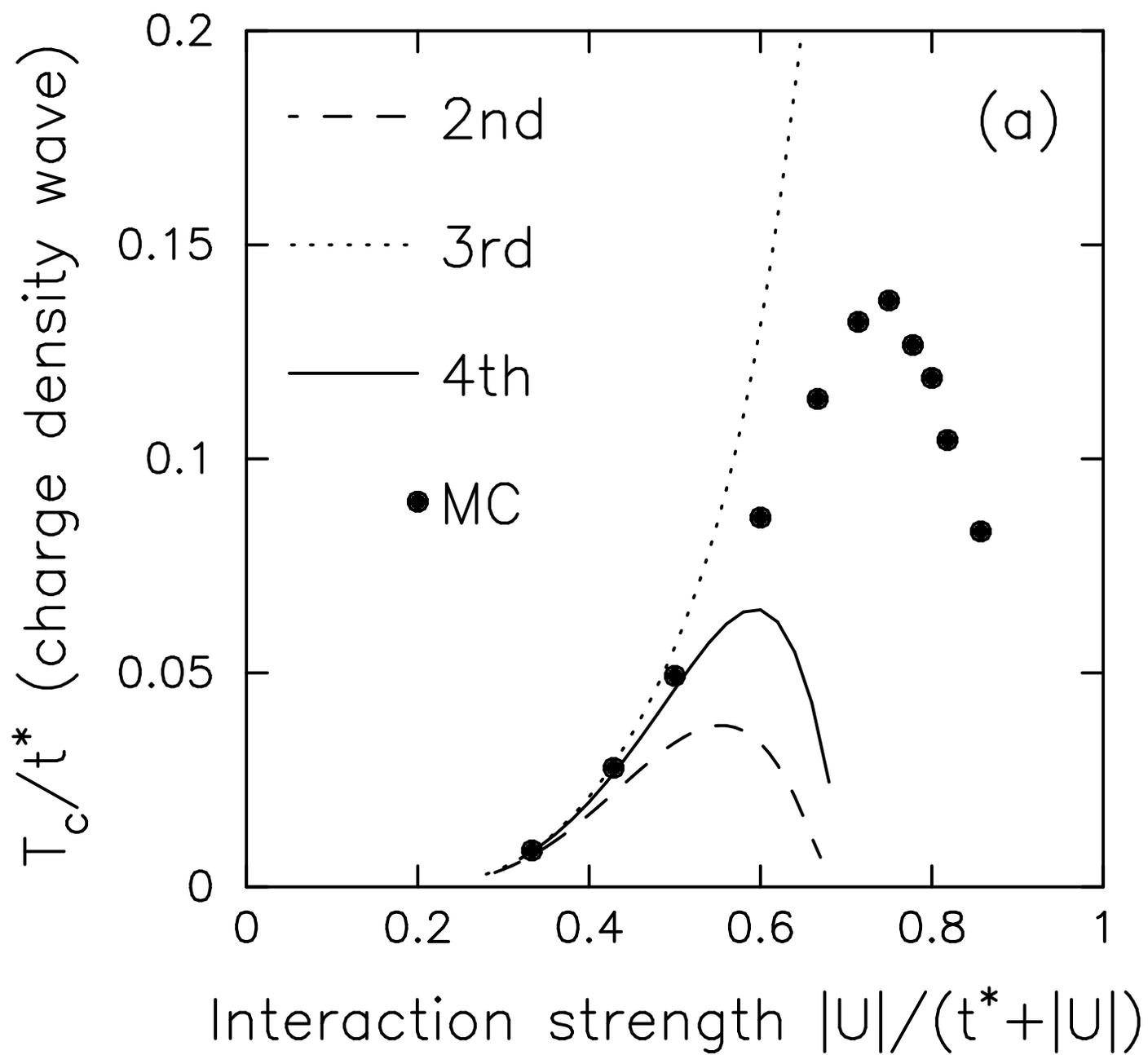
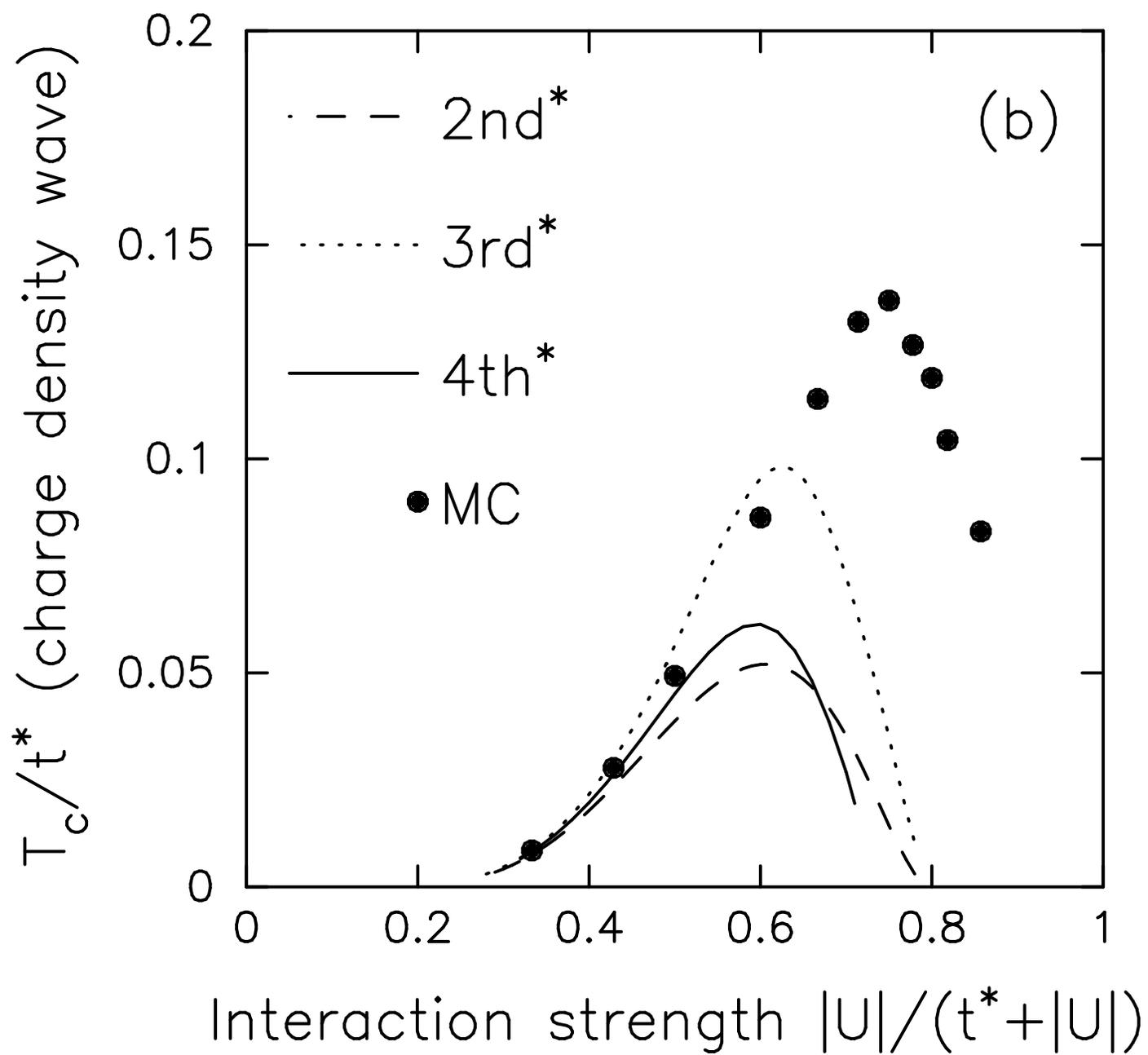
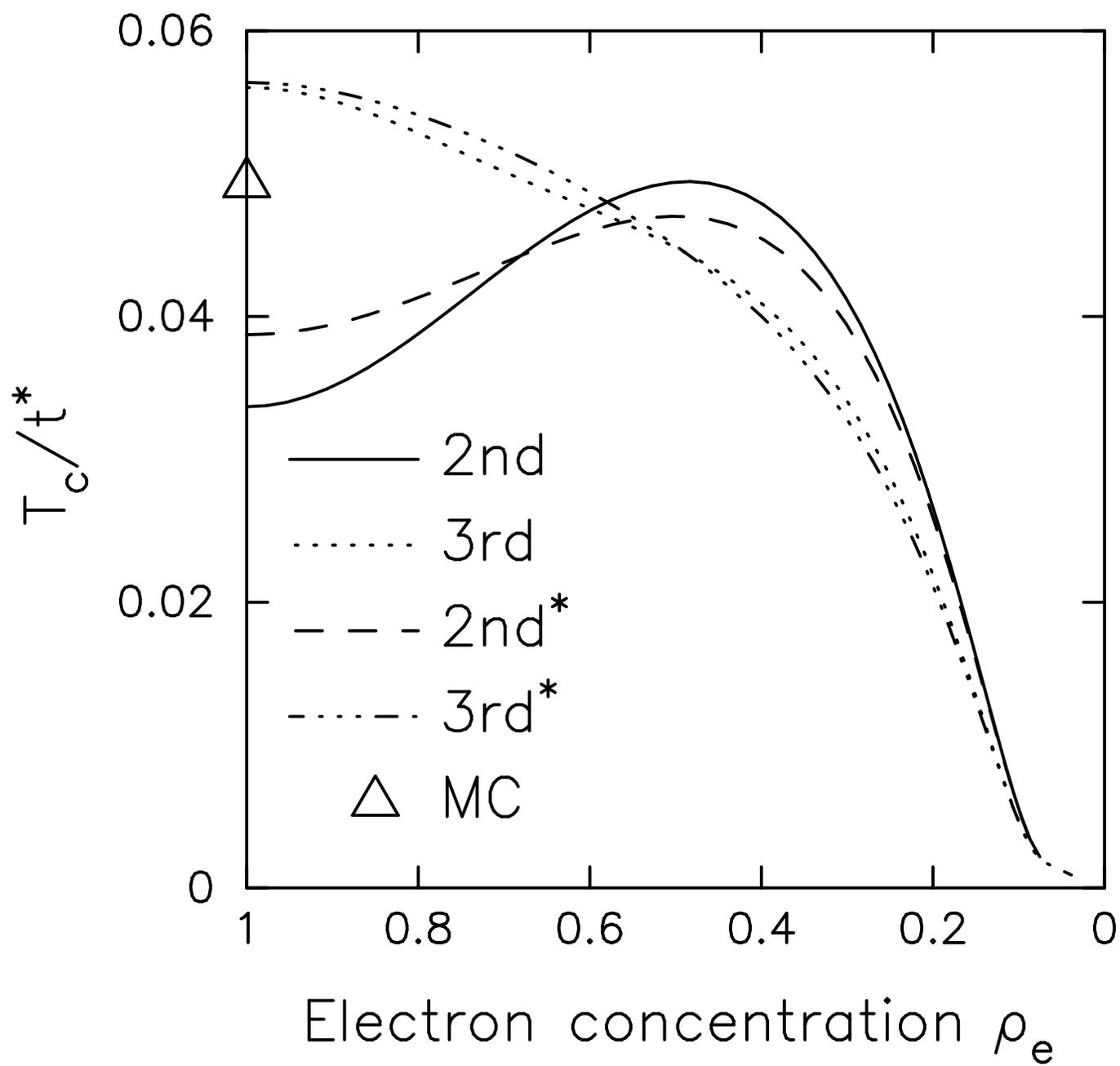
\end{document}